\documentclass[12pt]{iopart}
\usepackage{iopams}
\usepackage{graphicx}
\usepackage{amssymb,amsthm}

\def\re{\mathop{\rm Re}\nolimits}

\def\Tr{\mathop{\rm Tr}\nolimits}
\def\cltwo{\mathop{\rm Cl}_2\nolimits}

\begin{document}
\title[Penner matrix model with negative coupling constant]{Phase space and phase transitions in the Penner
                                                                                              matrix model with negative coupling constant}
\author{Gabriel \'Alvarez}
\address{Departamento de F\'{\i}sica Te\'orica II,
                               Facultad de Ciencias F\'{\i}sicas,
                               Universidad Complutense,
                               28040 Madrid, Spain}
\ead{galvarez@fis.ucm.es}
\author{Luis Mart\'{\i}nez Alonso}
\address{Departamento de F\'{\i}sica Te\'orica II,
                               Facultad de Ciencias F\'{\i}sicas,
                               Universidad Complutense,
                               28040 Madrid, Spain}
\ead{luism@fis.ucm.es}
\author{Elena Medina}
\address{Departamento de Matem\'aticas,
                      Facultad de Ciencias,
                      Universidad de C\'adiz,
                      11510 Puerto Real, Spain}
\ead{elena.medina@uca.es}
\begin{abstract}
The partition function of the Penner matrix model for both positive and negative values of the coupling constant
can be explicitly written  in terms of the Barnes $G$ function. In this paper we show that for negative values
of the coupling constant this partition function can also be represented as the product of an holomorphic matrix
integral by a nontrivial oscillatory function of  $n$.  We show that the planar limit of the free energy with {}'t~Hooft
sequences does not exist. Therefore we use a certain modification that uses Kuijlaars-McLaughlin sequences instead of
{}'t~Hooft sequences and leads to a well-defined planar free energy and to an associated two-dimensional phase space.
We describe the different configurations of complex saddle points of the holomorphic matrix integral
both to the left and to the right of the critical point, and interpret the phase transitions in terms
of processes of gap closing, eigenvalue tunneling, and Bose condensation.
\end{abstract}
\submitto{\JPA}
\maketitle
\section{Introduction}
The analysis of matrix models in the  large-$n$ limit is a lively area  of research for understanding relevant aspects of
the phase structure of  gauge theories~\cite{HO11,JA13,HO15,BUI16,AL16}.  In particular non-perturbative
effects attract a great deal of attention. They were first studied in the critical case with the double-scaling limit
method~\cite{DA91,DA93} but it was soon realized that they are also worth studying in matrix models away from
criticality~\cite{BO00,MA08,EY09,MAR09,MA08b,SC14,SC10,PA10,MA14}.

The  aim of the present paper is to analyze the phase space and the phase transitions of the Penner matrix model for
negative values of the coupling constant in the planar limit. Several of the results obtained reproduce aspects of
gauge theories which are usually described by unitary matrix models~\cite{HO11,HO15,BUI16,AL16,GR80,WA80}.  
The  discussion is based on  several important properties of (non-classical) generalized Laguerre polynomials which,
in particular, determine the exact solvability of the Penner matrix model and the complete explicit description
of its planar limit. 

The standard Hermitian Penner matrix model~\cite{PE88} is defined by the following formal partition function
\begin{equation}
	\label{mmp}
	\mathbf{ Z}_{n}(g)
	=
	\int_{H_n} \frac{\rmd X}{N_n}
	\exp\left[\frac{1}{g}\Tr\left( \sqrt{g} X+\log (1- \sqrt{g} X ) \right)\right], \quad g>0,
\end{equation}
where  the integration runs over the set $H_n$ of  $n\times n$ Hermitian matrices $X$, and the normalization
constant $N_n$ is
\begin{equation}
	\label{mmg}
	N_n
	=
	\int_{H_n} \rmd X \exp\left(-\frac{1}{2} \Tr X^2\right)=2^{n/2} \pi^{n^2/2}. 	
\end{equation}
The connection of the Penner model to continuum limits of theories of random discrete surfaces
and to the $c=1$ string theory is a consequence of  the application of the double-scaling
limit to the large-$n$ expansion of $\mathbf{ Z}_{n}(g)$ (see references~\cite{DI90,DI91} and the comment
following equation~(\ref{sas}) of appendix~A of the present paper). However, the double-scaling
limit is performed around a singular point of the large-$n$ expansion, which  exists only for negative values
of the coupling constant $g$~\cite{CP91,MA06}. This is our motivation to study the Penner model and its
large-$n$ limit in the region $g<0$.

The formal integral~(\ref{mmp}) was interpreted in references~\cite{PE88,MU98} as the limit of a sequence of matrix
integrals corresponding to truncations of the Taylor series of the exponent, and leads to a well-defined
topological expansion of the free energy. Remarkably~\cite{MU98}, the same asymptotic expansion can
be obtained from the eigenvalue integral
\begin{equation}
	\label{eig}
	Z_n(g)
	=
	C_n(g)\frac{1}{n!}
	\left(
		\prod_{i=1}^n \int_0^{\infty} \rmd x_i\,\rme^{-W_-(x_i)/g}
	\right)
	\Delta(\mathbf{x})^2,
\end{equation}	
where 
\begin{equation}
	\label{cn}
	C_n(g)
	=
	\frac{\rme^{n/g} g^{-n^2/2}}{(2\pi)^{n/2} \prod_{k=1}^{n-1} k!},
\end{equation}
$\Delta(\mathbf{x})$ is the Vandermonde determinant
\begin{equation}
	\label{van}
	\Delta(\mathbf{x}) = \prod_{j<k}(x_j-x_k),
\end{equation}
and
\begin{equation}
	\label{pot0}
 	W_-(x) = x-\ln x.
\end{equation}
In turn, the partition function $Z_n(g)$ is closely related to the family of  classical Laguerre polynomials 
$L^{(\alpha)}_n(z)$ ($\alpha>0$) (see for instance~\cite{PA10,DI90, DI91,CP91,MA06,DE02,AL14}),
and by applying the method of orthogonal polynomials the eigenvalue integral~(\ref{eig})
can be written in terms of the Barnes $G$ function as~\cite{PA10,AL14}
\begin{equation}
	\label{ane0}
	Z_n(g)
	=
	\frac{(\rme g) ^{n/g} g^{n^2/2}}{(2\pi)^{n/2}}
	\frac{G(1+n+\frac{1}{g} )}{G(1+\frac{1}{g})}, \quad g>0.
\end{equation}
Note that $|Z_n(g)|$ is well defined by the former expression not only for $g>0$ but also for all $g$ except $g=0$
and the zeros of the $G$ function in the denominator, i.e., for $g=-1/(k+1)$, $(k=1,2,3,\ldots)$.
Therefore, using the asymptotic expansions of the $G$ function quoted in appendix~A, we can find the asymptotic
expansion of the free energy
 \begin{equation}
 	\label{lef}
 	\mathcal{F}_n(g)  = - \frac{\ln|{Z}_n(-g)|}{n^2},\quad g>0.
 \end{equation}	

However, to study spectral aspects like the possible existence and qualitative behavior of the asymptotic eigenvalue density,
it is useful to have an eigenvalue integral representation of $Z_n(g)$ for $g<0$ (putting directly $g<0$ in equation~(\ref{eig})
leads to a divergent integral). This integral representation can be obtained by analytic continuation and
is the subject of section~\ref{sec:hooft}.

Section~\ref{sec:Barnes} is devoted  to the large-$n$ expansion of the free energy~(\ref{lef}).
The usual or {}'t~Hooft large-$n$ expansion is carried out with what in effect is a sequence of
coupling constants $g_n$ such that
\begin{equation}
	\label{tsec}
 	g_n n = t = {\rm constant}.
 \end{equation}
We show that for these {}'t~Hooft sequences the asymptotic expansion of the free energy $\mathcal{F}_n(g)$
is the sum of an oscillatory contribution and a perturbative contribution.
The perturbative contribution has essentially the same form as the large-$n$ expansion of  the Penner model
with positive coupling constant, and therefore it also  provides a generating function for
the virtual Euler characteristics of the spaces of Riemann surfaces with a finite number of punctures~\cite{PE88}.
However, we have not found the oscillatory contribution in the literature (for instance, it is missing in
references~\cite{CP91,MA06}). Because of the zeros of the Barnes $G(1-1/g)$ quoted above,
the free energy is singular at the value $t=1$ of the {}'t~Hooft parameter.

We also show that the oscillatory contribution does not converge in the planar limit for $t\neq 1$
with {}'t~Hooft sequences. In the context of generalized Laguerre polynomials, Kuijlaars and McLaughlin~\cite{KU01,KU04}
introduced coupling constant sequences, which hereafter we will call KM sequences, determined
by the two conditions
\begin{equation}
	\label{km}
	\lim_{n\rightarrow \infty}g_n n = t,
\end{equation}
(note that this condition is trivially satisfied by the {}'t~Hooft sequences),
and the existence of the limit
\begin{equation}
	\label{eq:l}
 	l = \lim_{n\rightarrow \infty} |\sin(\pi/g_n)|^{1/n}.
\end{equation}
In reference~\cite{AL15} we showed that the eigenvalue integral that gives the analytic continuation
of~(\ref{eig}) to negative values of $g$ (except for a prefactor discussed in section~\ref{sec:hooft}),
does not have a planar limit with {}'t~Hooft sequences in the strong-coupling region $1<t<\infty$,
but has a well defined planar limit with KM sequences, essentially because
the condition~(\ref{eq:l}) permits the handling of the nonvanishing oscillatory contribution.
In section~\ref{sec:Barnes} we also show that the planar limit of~(\ref{lef}) with {}'t~Hooft sequences also
does not exist in the weak-coupling region $0<t<1$, but does exit with KM sequences. This
result permits us to discuss the phase transitions in the $(t,l)$ plane and in particular the
phase transitions at $t=1$.

In section~\ref{sec:saddle} we use the eigenvalue integral to discuss the associated Coulomb
gas of eigenvalues and to describe the processes of  gap-closing and Bose condensation
of the asymptotic eigenvalue distribution. In section~\ref{sec:summary} we mention some matrix
models of physical interest which share similar properties and which can be studied with the same ideas.
The paper ends with two appendixes: in appendix~A we collect some results on the Barnes $G$ function,
and in appendix~B we apply these results to derive the celebrated topological expansion of the standard
Hermitian Penner model, which is compared in section~\ref{sec:Barnes} to the perturbative
contribution of the corresponding expansion for the free energy~(\ref{lef}).
\section{The partition function of the Penner model  for negative  values of the coupling constant \label{sec:hooft}}
\begin{figure}
    \begin{center}
        \includegraphics[width=8cm]{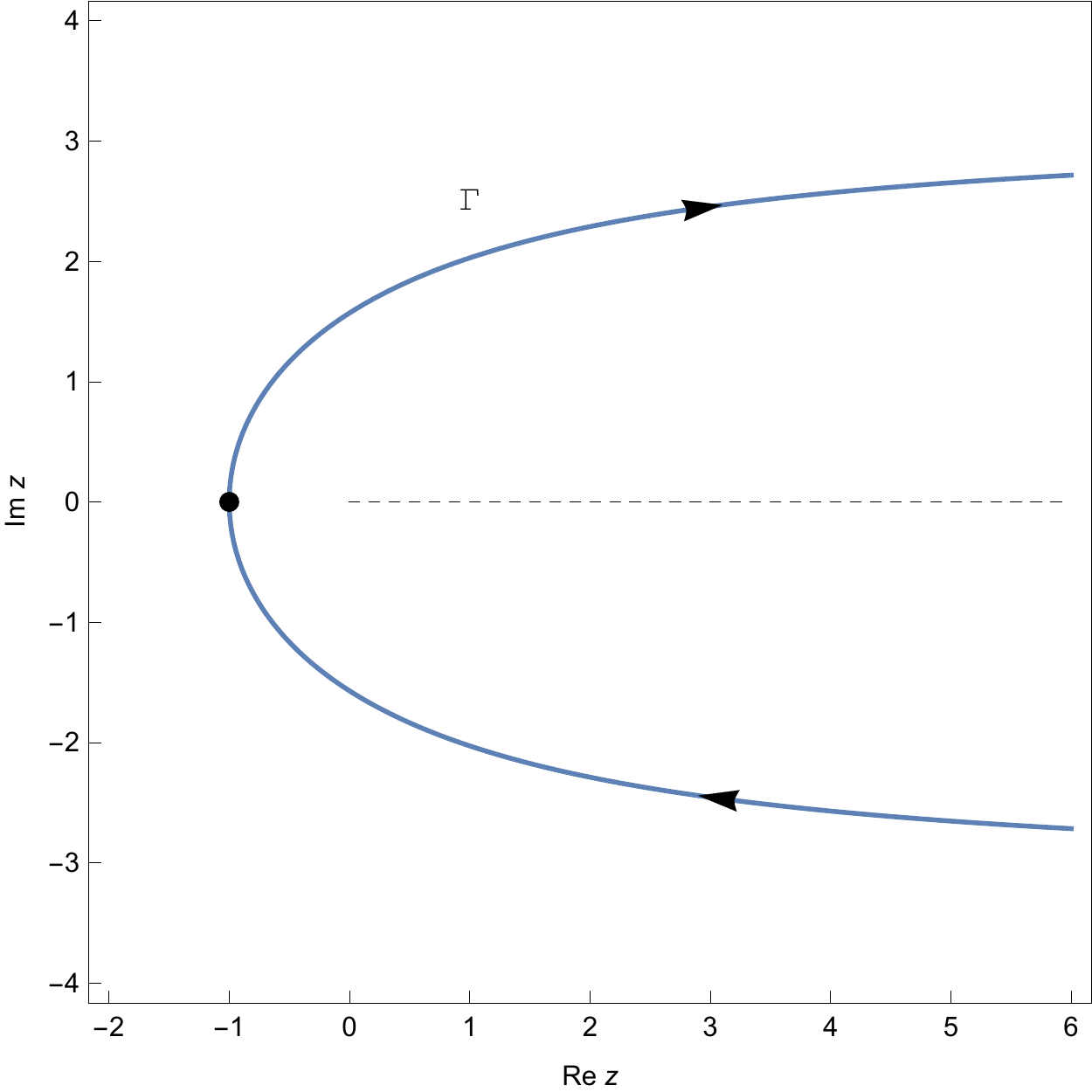}
    \end{center}
    \caption{Integration contour $\Gamma$ for the partition function~(\ref{eig1}).\label{fig:path}}
\end{figure}

\begin{figure}
    \begin{center}
        \includegraphics[width=8cm]{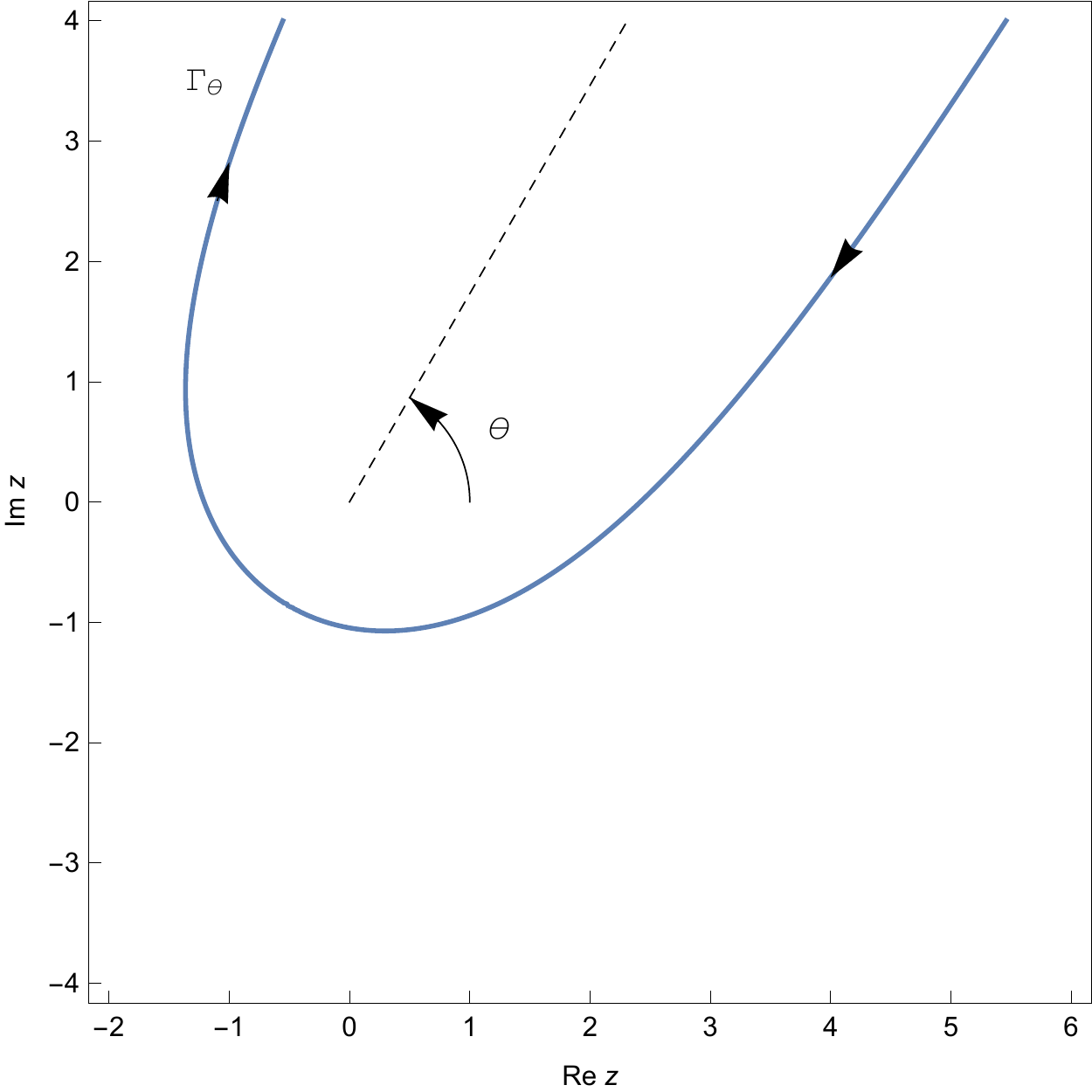}
    \end{center}
    \caption{Rotated integration contour $\Gamma_{\theta}$ for the analytic continuation of
                  the partition function~(\ref{eig1}).\label{fig:thetapath}}
\end{figure}

To perform the analytic continuation of the partition function~(\ref{eig}) to negative values of $g$
we first write it in the form 
\begin{equation}
	\label{eig1}
	Z_n(g)
	=
	\frac{C_n(g)}{(1-\rme^{2\pi\rmi/g})^{n}}\,\frac{1}{n!} \left(\prod_{i=1}^n \int_{\Gamma} 
	 \rmd z_i \,\rme^{-W_-(z_i)/g}\right)
	\Delta({\bf z})^2,
\end{equation}
where the contour $\Gamma$ is illustrated in figure~\ref{fig:path}.
Then we rotate counterclockwise the path $\Gamma$ to the path $\Gamma_{\theta}$ illustrated in
figure~\ref{fig:thetapath}, while simultaneously change the determination of $W_-(z)$ to
\begin{equation}
	\label{cw}
	W_\theta(z) = z-\log_{\theta} z,
\end{equation}
where $\log_{\theta} z= \ln|z|+\rmi \arg z$ with $\theta \leq \arg z< \theta+2\pi$. 
With these integration contour $\Gamma_{\theta}$ and determination $W_\theta(z)$, the rotated integral converges
in the half-plane $\theta-\pi/2<\arg g<\theta+\pi/2$, and that the rotated and unrotated integrals are equal
in $\theta<\arg g<\pi/2$ and $\theta-\pi/2<\arg g<0$.

Thus, setting $\theta=\pi$, we find that the analytic continuation $\mathcal{Z}_n(g)=Z_n^{({\rm cont})}(-g)$
for $g>0$ reads
\begin{equation}
	\label{eig2}
	\mathcal{Z}_n(g)	
	=
	\frac{C_n(-g)}{(1-\rme^{-2\pi\rmi/g})^{n}}\frac{1}{n!}
	\left(\prod_{i=1}^n\int_{\Gamma_{\pi}}\rmd z_i\,\rme^{(z_i-\log_{\pi} z_i)/g}\right)\Delta({\bf z})^2.
\end{equation}
Finally we change variables $z_i \to -z_i$, and taking into account that $\log_{\pi} (-z)=\log_0 z+\rmi \pi$
it follows that
\begin{equation} 
	\label{eig3}
	\mathcal{Z}_n(g)	
	=
	\frac{C_n(-g) (-1)^n\rme^{-\rmi n\frac{\pi}{g}}}{(1-\rme^{-2\pi\rmi/g})^{n}}
	\frac{1}{n!}\left(\prod_{i=1}^n\int_{\Gamma}\rmd z_i \rme^{- (z_i+\log z_i)/g}\right)\Delta({\bf z})^2,
\end{equation}
or, equivalently,
\begin{equation}
	\label{rel0}
	\mathcal{Z}_n(g) = C_n(-g) (-2\rmi)^{-n} (\sin( \pi/g))^{-n}\, \mathcal{Z}^{(0)}_n(g),
\end{equation}
where $\mathcal{Z}^{(0)}_n(g)$ is the holomorphic matrix integral~\cite{AL15} 
 \begin{equation}
 	\label{rel0h}
	\mathcal{Z}^{(0)}_n(g)
	=
	\frac{1}{n!}
	\left(
		\prod_{i=1}^n \int_{\Gamma} \rmd z_i \,\rme^{-W_+(z_i)/g}
	\right)
 	\Delta(\mathbf{z})^2,
\end{equation}
and
\begin{equation}
	\label{pee}
	W_+(z) = z+\log z.
\end{equation}
We stress that the partition function $\mathcal{Z}_n(g)$ given by equation~(\ref{eig2}) in effect defines
the Penner model for negative values of the coupling constant. Note in particular
the factorization into a prefactor (the first fraction) and the eigenvalue integral $\mathcal{Z}^{(0)}_n(g)$
studied in reference~\cite{AL15}.

As a check of this analytic continuation process, we recall that using generalized Laguerre polynomials
$L^{(-1/g)}_n(z /g)$~\cite{BE80,DI95} it was proved in reference~\cite{AL15} that  $\mathcal{Z}^{(0)}_n(g)$
can be expressed in terms of the Barnes function as
\begin{equation}
	\label{znp00}
\mathcal{Z}^{(0)}_n(g)=g^{n(n-\frac{1}{g})}(1-\rme^{-2\pi\rmi/g})^n
	          \frac{G(1+n)G\left(1+n-\frac{1}{g} \right)}{G\left(1-\frac{1}{g} \right)}.
\end{equation}
Hence, since
\begin{equation}
	\label{c}
	C_n(-g)=\frac{\rme^{-n/g}(-g)^{-n^2/2}}{(2\pi)^{n/2}G(1+n)},
\end{equation}
we finally get
\begin{equation}
	\label{ane10}
	\mathcal{Z}_n(g)
	=
	\frac{ (-\rme g)^{-n/g} (-g)^{n^2/2}}{(2\pi)^{n/2}}
	\frac{G(1+n-\frac{1}{g} )}{G(1-\frac{1}{g})}, \quad g>0,
\end{equation}
which is the  same expression obtained by replacing $g\to -g$ in equation~(\ref{ane0}).

Hereafter we will shorten the full qualification ``Penner model with negative coupling constant'' to simply
``Penner model'' when we refer to the model specified by the partition function $\mathcal{Z}_n(g)$.
\section{The large-$n$ limit  of the free energy \label{sec:Barnes}}
\subsection{The large-$n$ limit with {}'t~Hooft sequences}
Let us analyze the large-$n$ limit of the partition function $\mathcal{Z}_n(g)$ with {}'t~Hooft sequences $g_n=t/n$.
From~(\ref{ane10}) and applying  the reflection formula~(\ref{bneg}) to $\ln\left|  G\left(1+n-1/g\right) \right|$ and
to $\ln \left|G\left(1-1/g\right) \right|$  we have
\begin{eqnarray}
\nonumber 	\mathcal{F}_n(g_n) & = & - \frac{\ln|\mathcal{Z}_n(g_n)|}{n^2}\\	\label{ane2mn}
	                                     & = & -\frac{1}{2n}\ln\left(\frac{\pi}{2}\right)+\left(\frac{1}{ng_n}-\frac{1}{2}\right)\ln g_n
	                                              + \ln\left|\sin\left(\frac{\pi}{g_n}\right)\right|^{1/n}\nonumber\\
                                             &   & {}+\frac{1}{ng_n} -\frac{1}{n^2}\ln\left|  G\left(1-n+\frac{1}{g_n}\right) \right|
                                                                            +\frac{1}{n^2}\ln \left|G\left(1+\frac{1}{g_n}\right) \right|.
\end{eqnarray}
Since $1+1/g_n >0$, the term $\ln |G(1+1/g_n)|$ in~(\ref{ane2mn}) has a Stirling expansion
of the type~(\ref{id1}). However, the sign of the argument in
$\ln |G(1-n+1/g_n)|$ depends on the value of $t$. Indeed,  for $0<t<1$ and large $n$ we have
\begin{equation}
	\label{tm}
	\ln \Big |G\left(1-n+\frac{1}{g_n}\right)\Big |= \ln\Big |G\left(1+x \right)\Big |, \quad x=n \left(\frac{1}{t}-1\right)\rightarrow +\infty,
\end{equation}
so  that   this function has  also a Stirling expansion of the type~(\ref{id1}).  However, for $1<t<\infty$ and large $n$   we have
\begin{equation}
	\label{tm2}
	\ln \Big |G\left(1-n+\frac{1}{g_n}\right)\Big| = \ln\Big| G\left(1-x \right)\Big |, \quad x=n \left(1-\frac{1}{t}\right)\rightarrow +\infty,
\end{equation}
and we have to use the expansion~(\ref{nxb}) of Appendix A. Therefore, in this region
the free energy can be decomposed into a sum of an oscillatory contribution and a perturbative contribution
  \begin{equation}\label{des}
 \mathcal{F}_n(t)=\mathcal{F}^{(\rm osc)}_n(t)+\mathcal{F}^{(\rm per)}_n(t),
 \end{equation}
where
\begin{equation}
	\label{nopf}
	\mathcal{F}^{(\rm osc)}_n(t)
	=
	\left\{\begin{array}{ll}\ln \left|2\sin( \pi n/t) \right|^{1/n}, & \mbox{for $0<t<1$,} \\
	\frac{1}{t}\ln \left|2\sin(\pi n/t) \right|^{1/n}+\frac{1}{2\pi n^2}{\rm Cl}_2\left(2\pi n/t\right),
	& \mbox{for $1<t<\infty$,}\end{array}\right.
\end{equation}
and in both cases, i.e., for $t\neq 1$, the perturbative contribution is
\begin{eqnarray}\label{top3}                                                                
\nonumber  \mathcal{F}_n^{{\rm (per)}}(t) &\approx& -\left(\frac{(t-1)^2}{2t^2} \ln |1-t|-\frac{3}{4}+\frac{1}{2 t}\right)
                                                                +\frac{1}{12 n^2}\ln|1-t|
                                         \\ 
                                                            & -&\sum_{k=2}^\infty \frac{B_{2k}}{2k(2k-2)} n^{-2k}t^{2k-2}
                                                                                             \left( (1-t)^{2-2k}-1 \right), \quad n\rightarrow \infty.                                                                           
\end{eqnarray}
The perturbative contribution  $ \mathcal{F}_n^{{\rm (per)}}(t) $ coincides with the standard large-$n$ expansion for the
Penner model with negative coupling constant~\cite{CP91,MA06}. Comparing the  expression~(\ref{top3})  with the
topological expansion (\ref{top})--(\ref{chii}) of the Penner model with positive coupling constant, it is clear that 
$\mathcal{F}_n^{{\rm (per)}}(t)$ also represents a  generating function for the virtual Euler characteristics.
To the best of our knowledge, the oscillatory contribution $\mathcal{F}^{(\rm osc)}_n(t)$ has not been considered
in the literature. Note also that the value $t=1$ is a critical value for both contributions.
\subsection{The planar limit}
The planar limit of the free energy 
 \begin{equation}
\label{fee}
	\mathcal{F}=  \lim_{n\rightarrow \infty} \mathcal{F}_n,
\end{equation}
is not well-defined with the  standard {}'t~Hooft sequences~(\ref{tsec}).
Indeed, from the expression (\ref{nopf}) of $\mathcal{F}^{(\rm osc)}_n(t)$
it is clear that the existence of  $\mathcal{F}$ requires  the existence of the limit $l$ defined in equation~(\ref{eq:l}).

Thus, instead of restricting to sequences satisfying the  {}'t~Hooft coupling, we use
KM sequences.  From equations (\ref{des})--(\ref{top3}) it follows that  for $t\neq 1$
 \begin{equation}
 	\label{nopfp}
	\mathcal{F}= \left\{\begin{array}{cc}\ln l -\frac{(t-1)^2}{2t^2} \ln |t-1|+\frac{3}{4}-
	\frac{1}{2 t},
 & \mbox{for $0<t<1$,} \\ &  \\ \frac{1}{t}\,\ln l -\frac{(t-1)^2}{2t^2} \ln |t-1|+\frac{3}{4}-
	\frac{1}{2 t},
&  \mbox{for $1<t<\infty$.} \end{array}\right.
\end{equation}
It is illustrative to clarify the origin of the different terms in equation~(\ref{nopfp}) with respect to the factors
in the equation~(\ref{rel0}). From equation~(\ref{c}) and taking into account  that
\begin{equation}
	\label{asb}
	\frac{1}{n^2}\ln |G(1+n)|
	=
	\frac{1}{2}\ln n-\frac{3}{4}+ O(1/n),\quad n\rightarrow \infty,
\end{equation}
it follows that
\begin{equation}
	\label{pers012}
 	\mathcal{F} =  \ln l+\frac{1}{t}+\frac{1}{2}\ln t+\frac{3}{4}+ \mathcal{F}^{(0)},
\end{equation}
where $ \mathcal{F}^{(0)}$ is the planar free energy for $\mathcal{Z}^{(0)}_n$~\cite{AL15}
\begin{equation}
	\label{llz}
\fl  	\mathcal{F}^{(0)} = H(t-1)\left(\frac{1}{t}-1 \right) \ln l
 				-\frac{1}{2}\ln t+\frac{3}{2}\left(\frac{t-1}{t}\right)
				- \frac{(t-1)^2}{2t^2} \ln |t-1|,
\end{equation}
where $H(x)$ is the Heaviside step function.  It is now clear that the term $\ln l$  of $\mathcal{F}$ in the
weak-coupling case comes only from the factor $|\sin(\pi/g)|^{-n}$ in (\ref{rel0}), while the term $(\ln l )/t$
of  $\mathcal{F}$  in the strong-coupling case  is a result of the contributions from the two factors
$|\sin(\pi/g)|^{-n}$ and $|\mathcal{Z}^{(0)}_n|$ in~(\ref{rel0}).
\subsection{The phase space of the Penner model for negative values of the coupling constant}
\begin{figure}
    \begin{center}
        \includegraphics[width=8cm]{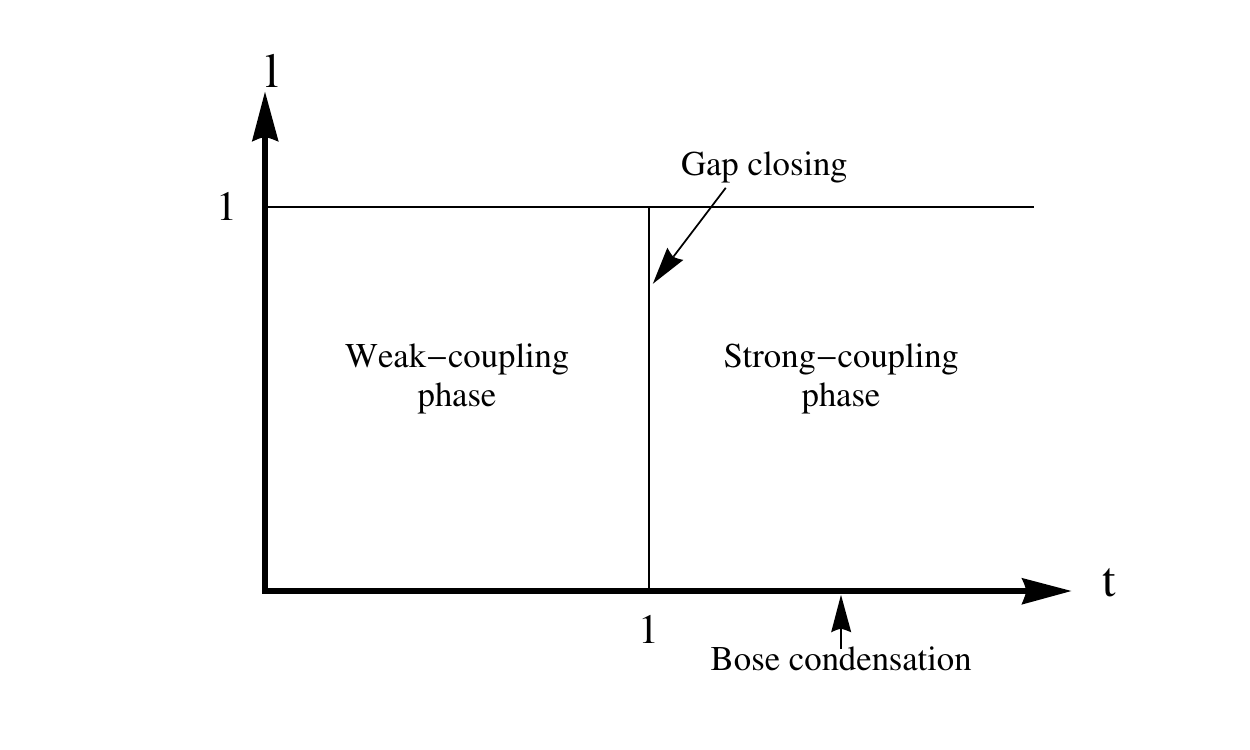}
    \end{center}
    \caption{Phase space of the non-Hermitian Penner model with negative coupling constant
                  for KM sequences.\label{fig:path4}}
\end{figure}

The $l$-dependence of the planar free energy means that  from the point of view of KM sequences, the phase space
of the Penner model in the planar limit  is the set  
\begin{equation}\label{pes}
 	\mathcal{P} = \{(t,l)\in \mathbb{R}^2\, | \, 0<t<\infty,\; 0\leq l\leq 1\}.
\end{equation}
In this way KM sequences reveal a fine phase space structure (see figure~\ref{fig:path4}), where both the weak- and
strong-coupling phases depend on the additional parameter $l$. The points  $(t,l)=(1,l)$  for $0<l<1$ represent
first-order phase transitions
\begin{equation}
	\label{first}
 	\mathcal{F}\Big|_{t=1-0}
	=
	\mathcal{F}\Big|_{t=1+0},\quad \frac{\partial  \mathcal{F}}{\partial t}\Big|_{t=1-0}
	=
	\frac{\partial  \mathcal{F}}{\partial t}\Big|_{t=1+0}+\ln l,
\end{equation}
while the point $(t,l)=(1,1)$ represents a continuous  phase transition in which $ \mathcal{F}$ and $\partial\mathcal{F}/\partial t$
are continuous at $t=1$ but  $\partial^2  \mathcal{F}/\partial t^2$ diverges. Furthermore, the line $l=0$ represents a
singular phase with  infinite free energy. 

It is possible to characterize wide classes of KM sequences for all $0<t<\infty$ and $0\leq l\leq 1$,
as for example the one-parameter family
\begin{equation}
	g_n = \frac{1}{[n/t]+c\,l^n}, \quad c\neq 0,
\end{equation}
where $[x]$ denotes the integer part of $x$. They also arise as subsequences of {}'t~Hooft sequences $g_n=t/n$.
For example, given an irreducible fraction $t=p/q$ $(p>1)$, the subsequences $g_{np}$ and $g_{np+1}$
of the {}'t~Hooft sequence are KM sequences with $l=0$ and $l=1$, respectively.  For irrational  $t$ 
we have that  the sequence $\{n/t\}$, where $\{x\}$  denotes  the  fractional part of $x$ is a dense subset
of the interval $[0,1]$, so that all subsequences of the {}'t~Hooft sequence $g_n=t/n$ such that $\{k_n/t\}\rightarrow x$
for some $0<x<1$ are KM sequences with $l=1$. The  value  $l=1$  represents the generic case of KM
sequences (see remark~1.3 in~\cite{KU04}). 

A deeper understanding of the phase space of the Penner model is obtained from the analysis of the asymptotic
eigenvalue (saddle point) distribution. 
\section{Large-$n$ saddle points and Coulomb gas}\label{sec:saddle}
The  KM sequences originated in the theory of large-$n$ asymptotics of zeros of generalized Laguerre
polynomials~\cite{KU01,KU04}
\begin{equation}
	\label{lagp}
	L^{(\alpha)}_n(z)
	=
	\sum_{k=0}^n\left(\begin{array}{c}n+\alpha \\n-k\end{array}\right)\frac{(-z)^k}{k!}, \quad \alpha<0.
\end{equation}
In~\cite{AL15} we showed how these zeros determine  the saddle points of the holomorphic matrix integral 
$\mathcal{Z}^{(0)}_n(g)$  for KM sequences in the strong-coupling case.  In order to describe the phase
transitions at $t=1$, we will extend here the analysis of~\cite{AL15} by considering 
the weak-coupling case too.
\subsection{Complex saddle points}
The saddle point equations for $\mathcal{Z}^{(0)}_n(g)$  are
\begin{equation}
	\label{sa0}
	\frac{1}{g_n}\left(1+ \frac{1}{z_i^{(n)}}\right)
	+
	\sum_{j\neq i}\frac{2}{z_j^{(n)}-z_i^{(n)}}
	=
	0,\quad i=1,\ldots,n,
\end{equation}
and their solutions are given by~\cite{AL15}
\begin{equation}
	\label{sopl}
	z_i^{(n)} = g_n\,l^{(\alpha_n,n)}_i,\quad i=1,\ldots,n,
\end{equation}
where  $l^{(\alpha_n,n)}_i$ are the zeros of the Laguerre polynomials $L^{(\alpha_n)}_n(z)$ with
\begin{equation}
	\label{zeroo}
 	\alpha_n=-1- \frac{1}{g_n}.
 \end{equation}
 
Let us denote by $\rho(z)$ the asymptotic eigenvalue (saddle point) distribution for the Penner model and by
$\rho_L(z)$ the asymptotic  zero distribution of the scaled Laguerre polynomials  $L^{(\alpha_n)}_n(nz)$.
The form of $\rho_L(z)$  has been completely characterized  in the large-$n$ limit 
\begin{equation}
	\label{clag}
	n\rightarrow \infty,\quad \frac{\alpha_n}{n}\rightarrow A = \mbox{fixed},
\end{equation}
for all real values of $A$ in references~\cite{KU01,KU04,MA01,DI11}. Using~(\ref{sopl}) and~(\ref{zeroo})    
we can immediately translate the  properties of $\rho_L(z)$  into properties  of $\rho(z)$, since
\begin{equation}
	\label{tra}
  	t = -\frac{1}{A}, \quad \rho(z)=\frac{1}{t}\rho_L\left(\frac{z}{t}\right).
\end{equation}

It turns out that the zeros of $L^{(\alpha_n)}_n(nz)$ cluster along certain curves $\gamma_L$
in the complex plane. More concretely, if we denote 
 \begin{equation}
 	\label{end}
	a_{\pm} = A+2\pm 2\sqrt{A+1},
\end{equation}
\begin{enumerate} 
 	\item For $-\infty< A<-1$ (the weak-coupling region $0<t<1$)  the curve $\gamma_L$
	         is a simple open arc with endpoints $a_-$ and $a_+=\overline{a}_-$ symmetric
	         with respect to $ \mathbb{R}$, which as $A\rightarrow -1$ closes
	         and becomes the Szeg\H{o} curve
 		\begin{equation}
			\label{zego}
			|z\,\rme^{1-z}|=1,\quad |z|\leq 1.
 		\end{equation}
  		This  process is shown in figure~\ref{fig:path1}. The corresponding zero density is
 		\begin{equation}
			\label{cl0}
			\rho_L(z) = \frac{1}{2\pi}\left|\frac{\sqrt{(z-a_-)(z-a_+)}}{z}\right|.
 		\end{equation}
	 \item For $-1< A<0$ (the strong-coupling region $1<t<\infty$), and if the limit 
 		\begin{equation}
			\label{eq:llll}
 			l=\lim_{n\rightarrow \infty} |\sin(\alpha_n \pi)|^{1/n}= \lim_{n\rightarrow \infty} |\sin(\pi/g_n)|^{1/n},
 		\end{equation}
		exists, then $\gamma_L$ is of the form
 		\begin{equation}
			\label{ge2}
			\gamma_L=C_l \cup [a_-,a_+],
		\end{equation}
 		where 
 		\begin{enumerate}
 			\item For $l\neq 0$ the curve $C_l \subset \mathbb{C}\setminus (\{0\}\cup [a_-,+\infty))$
				is a simple closed curve encircling $0$ once, which is determined  by the implicit equation
				\begin{equation}
					\label{cl1}
 					{\rm Re}\,\int_{a_-}^z \frac{\sqrt{(z'-a_-)(z'-a_+)}}{z'}{\rm d}z'=-\log l.
 				\end{equation}
 				The corresponding zero density is
 				\begin{equation}
					\label{cl2}
 					\rho_L(z)=\frac{1}{2\pi}\left|\frac{\sqrt{(z-a_-)(z-a_+)}}{z}\right|.
				\end{equation}
  			\item For $l=0$ 
				\begin{equation}
					\label{ge20}
					\gamma_L=\{0\}\cup [a_-,a_+],
				\end{equation}
				and  the zero density is 
 				\begin{equation}
					\label{cl3}
					\rho_L(z)
					=
					A\, \delta(z)+\frac{1}{2\pi}\left|\frac{\sqrt{(x-a_-)(x-a_+)}}{x}\right|\chi_{[a_-,a_+]},
 				\end{equation}
		\end{enumerate}
		where $\chi_{[a_-,a_+]}$ is the characteristic function of the real interval $[a_-,a_+]$.
 		The filling fraction of the zero density on $C_l$ is equal to $|A|$.  The  value  $l=1$ is
		the generic case  (see remark~1.3 in~reference \cite{KU04}), and it follows from
		equation~(\ref{cl1}) that it is the only situation in which  the loop $C_l$ and the interval $[a_-,a_+]$
		intersect (at  the point $a_-$). It should be noticed that the form of $\gamma_L$  depends not only
		on the value of $A$ but  also on $l$.  
\end{enumerate}
\subsection{Coulomb gas, gap closing, eigenvalue tunneling and  Bose condensation}
In this section we use the electrostatic interpretation wherein the eigenvalue density $\rho(z)$ is thought of as a 
unit normalized positive charge density for a Coulomb gas in the external electrostatic potential 
\begin{equation}
	\label{uve}
	V(z)	= x+\ln |z|,\quad (x=\re z).
\end{equation}
The  electrostatic energy of the Coulomb gas,
\begin{equation}
	\label{efe}
	\mathcal{E}(t)=\frac{1}{t}\int_{\gamma}  V(z) \rho(z)|{\rmd}z|
	-
	\int_{\gamma} |\rmd z| \int_{\gamma} |\rmd z'| \ln |z-z'| \rho(z) \rho(z'),
\end{equation}
is given~\cite{AL15} by
\begin{equation}
	\label{ee}
	\mathcal{E}(t)=-\frac{1}{2}\ln t-\frac{(t-1)^2}{2t^2} \ln |t-1|+\frac{3}{2}\left(1-\frac{1}{t}\right).
\end{equation}
We remark that $\mathcal{E}(t)$ is independent of $l$.
\begin{figure}
    \begin{center}
        \includegraphics[width=8cm]{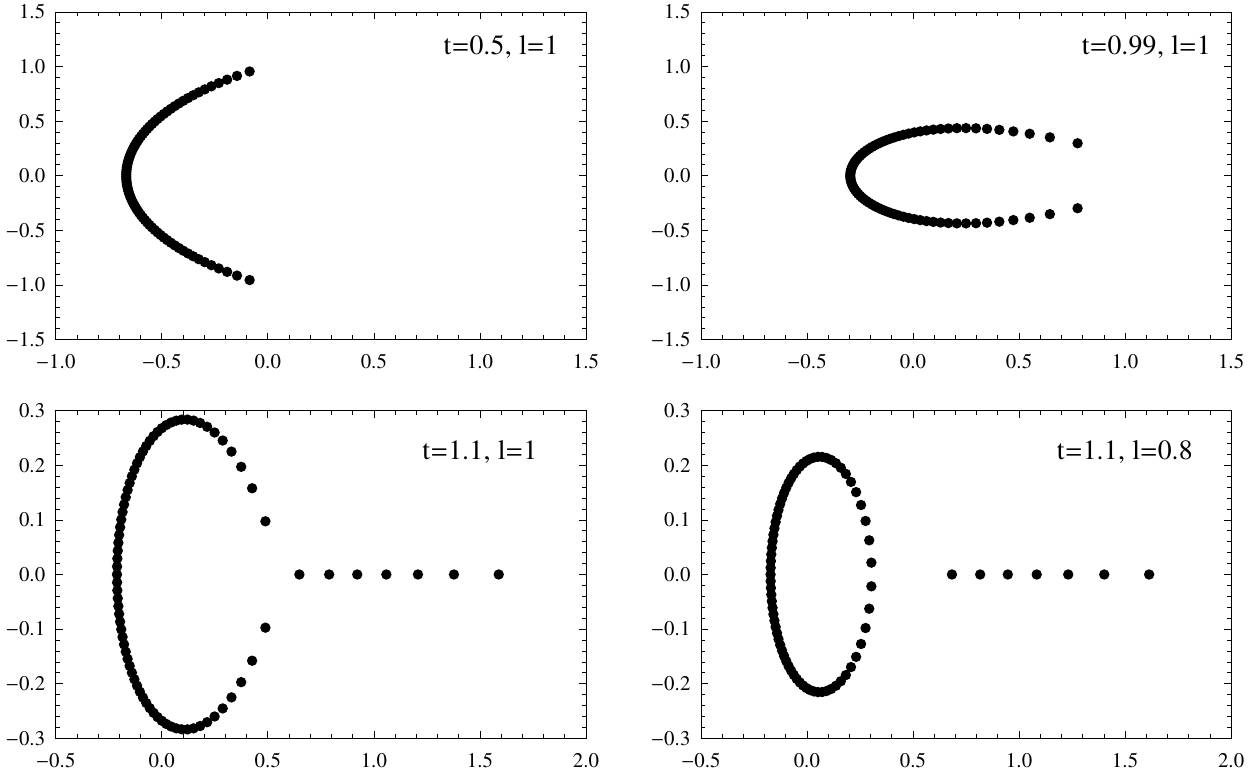}
    \end{center}
    \caption{Eigenvalue distribution in the gap-closing  phase transition for $n=80$.\label{fig:path1}}
\end{figure}

The  form  of the support of  $\rho(z)$ is independent of $l$ in the weak-coupling phase, but it depends
on $l$ in the strong-coupling phase. Thus for $1<t<\infty$ the support  of $\rho(z)$ consists of two pieces:
an $l$-dependent closed loop $\gamma_1$ around the origin with filling fraction $1/t$, and  an $l$-independent interval
$\gamma_2=[ta_-,ta_+]$ on the positive real axis with filling fraction $1-1/t$.  

The points of the weak-coupling phase are electrostatic stable equilibrium configurations of $\rho(z)$. However, points
of the strong-coupling phase represent stable electrostatic equilibrium configurations only for $l=1$~\cite{AL15}.
Indeed, for strong coupling the effective  potential of the charge distribution corresponding to $\rho(z)$
\begin{equation}
	\label{loge0}
	V_{\rm eff}(z) = V(z) - 2 t \int_{\gamma}  \ln |z-z'| \rho(z') |\rmd z'|,
\end{equation}
is constant on the two pieces of the support of $\gamma$~\cite{AL15}, but with  values 
\begin{equation}
	\label{une}
	V_{\rm eff}\big |_{\gamma_1} =  - t\ln l+V_{\rm eff}\big |_{\gamma_2},
	\quad
	V_{\rm eff}\big |_{\gamma_2}=(2t-1) -t \ln t - (t-1) \ln (t-1),
\end{equation}
which coincide  for $l=1$ only.

\begin{figure}
    \begin{center}
        \includegraphics[width=8cm]{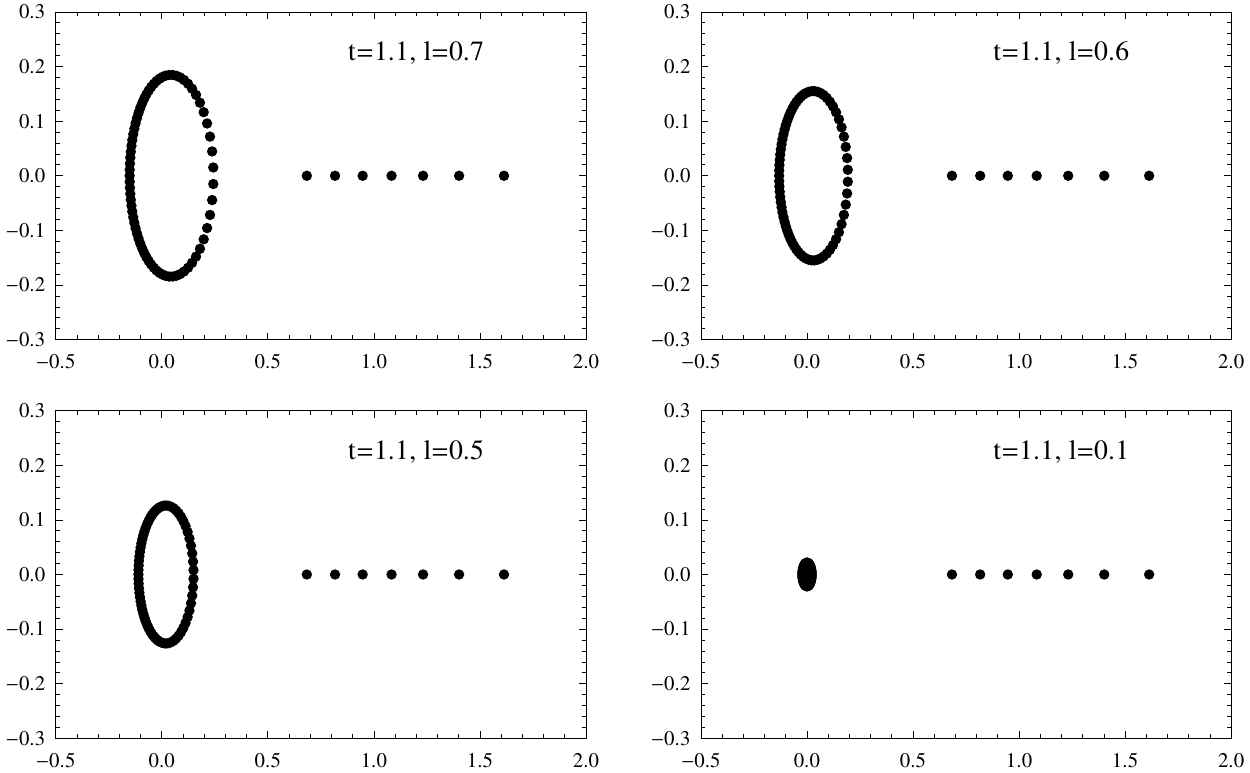}
    \end{center}
    \caption{Eigenvalue distribution in  Bose condensation  for $n=80$.\label{fig:bose}}
\end{figure}

Finally, note the following two types of phase transitions:
\begin{enumerate}
	\item In the weak-coupling phase the limit $t\rightarrow 1$  and constant $l\neq 0$ represents
	         a gap-closing (confining)  transition in which the open arc formed by the support of $\rho(z)$
	         closes and  becomes the Szeg\H{o} curve (figure~\ref{fig:path1}). In the strong-coupling phase
	         this limit  represents a gap-opening (deconfining) transition mediated by  an eigenvalue tunneling process
	         in which the charge $1-1/t$ located on the interval $\gamma_2$ tunnels  to the higher-potential points
	         of the loop $\gamma_1$.
	\item As  $l\rightarrow 0$ from the strong-coupling phase  with constant $t$ the loop $\gamma_1$ of the support of
		$\rho(z)$ shrinks to the point $z=0$ forming a condensate of charge $1/t$ (figure~\ref{fig:bose}). This represents
		a process of Bose condensation in which  the electrostatic energy $\mathcal{E}$ (\ref{ee}) remains
		finite and constant. However, the planar free energy diverges ($\mathcal{F}\rightarrow -\infty$)
		as a consequence of the contribution to $\mathcal{F}$ of the  oscillatory factor in~(\ref{rel0}). 
\end{enumerate}
\section{Outlook\label{sec:summary}}
There are several relevant non-Hermitian matrix models which have a phase space in the large-$n$ limit with properties
similar to the  Penner model with negative coupling constant.  We will briefly mention some examples.

Multi-Penner models of the form 
\begin{equation}
	\label{mp}
	W(z) = -\sum_{i=1}^k \mu_i \log (z-q_i),
\end{equation}
have been introduced to characterize the correlation functions of the $d=2$ conformal $A_1$
Toda field theory~\cite{SC10,DI09}.  The simplest case is the double Penner model~\cite{SC10}
\begin{equation}
	\label{jac}
	W(z) = -\mu_+\log (1-z)-\mu_- \log(z+1),
\end{equation}
which is connected to the theory of Jacobi polynomials
\begin{equation}
	\label{j1}
	P^{(\alpha,\beta)}_n(z)
	=
	2^{-n}
	\sum_{k=0}^n
		\left(\begin{array}{c}n+\alpha \\n-k\end{array}\right)
		\left(\begin{array}{c}n+\beta \\k\end{array}\right)(z-1)^k(z+1)^{n-k},
\end{equation}
with
\begin{equation}\label{j2}
 \alpha=\frac{\mu_+}{g},\quad \beta=\frac{\mu_-}{g}.
\end{equation}
The Hermitian matrix model  corresponds to the classical Jacobi polynomials with $\alpha,\beta>-1$.
Its large-$n$ limit with {}'t~Hooft sequences is well defined, and the eigenvalue distribution is determined
by the asymptotic zero distribution of the Jacobi polynomials on the real interval $[-1,1]$. However, there are
non-classical cases for which the existence of the planar limit requires a special formulation based on KM
sequences~\cite{MA05}. For example, if we take a large-$n$ limit with sequences $n g_n\rightarrow t>0$ such that 
\begin{equation}
	\label{j3}
	n\rightarrow \infty,\quad \frac{\alpha_n}{n}\rightarrow A
	=
	\frac{\mu_+}{t},\quad \frac{\beta_n}{n}\rightarrow B
	=
	\frac{\mu_-}{t},
\end{equation}
where
\begin{equation}
	\label{j4}
	-1<A<0<B,
\end{equation}
or, equivalently,
\begin{equation}
	\label{j5}
\mu_+<0,\quad \mu_->0,\quad t>|\mu_+|.
\end{equation}
Then the existence of a well-defined asymptotic zero distribution requires the existence of the limit~\cite{MA05} 
\begin{equation}\label{j6}
l= \lim_{n\rightarrow \infty}|\sin( \pi A n)|^{1/n}.
\end{equation}
The corresponding  zero distribution of non-classical Jacobi polynomials exhibits  properties
like gap-closing processes~\cite{MA05},  similar to the generalized  Laguerre polynomials. Therefore it would
be interesting to analyze the non-Hermitian version of the double Penner model  following the scheme used
in the present paper for the Penner model.

Families of  Laguerre polynomials also appear in certain matrix models used to study the low-energy
limit in Quantum Chromodynamics, e.g., the chiral Gaussian Unitary matrix model 
(also called the Wishart-Laguerre ensemble)~\cite{VE05,AK16} 
\begin{equation}
	\label{ch1}
 	Z_n^{(N_f,\nu)}
	=
	\frac{1}{n!}\left( \prod_{i=1}^n \int_0^{\infty}
	 \rmd x_i \, x_i^{\nu}\rme^{-x_i}\prod_{f=1}^{N_f}(x_i+m_f^2)\right)\Delta(\mathbf{x})^2,
\end{equation}
where $N_f>0$ is the number of fermionic quark flavors, $m_f$ $(f=1,\ldots,N_f)$ are the corresponding
masses, and $\nu\geq 0$ is the topological charge of the physical sector under consideration.

This partition function can be determined in terms of Laguerre polynomials (see~\cite{AK16} and the references therein).
For example, it is clear that the reduced case with $m_f=0$ for all $f$ gives
\begin{equation}
	\label{ch1r}
 	\mathcal{Z}_n^{(N_f,\nu)}
	=
	\frac{1}{n!}\left( \prod_{i=1}^n \int_0^{\infty}
	 \rmd x_i \, x_i^{\nu+N_f}\rme^{-x_i}\right)\Delta(\mathbf{x})^2,
\end{equation}
which is directly associated to the family of Laguerre polynomials $L^{(\nu+N_f)}_n(x)$. Consequently,
non-Hermitian versions of~(\ref{ch1r}), like those arising from~(\ref{ch1})  with bosonic quarks
($N_f<0$)~\cite{SP06}, will involve families of generalized Laguerre polynomials and may exhibit
non-perturbative effects similar to the Penner model with negative coupling constant. 
\section*{Appendix A. The Barnes $G$ function}
This Appendix collects the properties that we need about the Barnes $G$ function~\cite{BA00,OL10,AD01}.
The Barnes $G$ function is the entire function defined by the canonical product
\begin{equation}
 	G(1+z) = (2\pi)^{z/2}
	               \rme^{-\frac{1}{2}(z+z^2(1+\gamma))}
	               \prod_{k=1}^\infty \left(1+\frac{z}{k}\right)^k \rme^{-z+z^2/2k}.
\end{equation}
The Stirling-like asymptotic expansion of $G(1+z)$ for $z=x$ positive and large is
\begin{eqnarray}
\ln G(1+x) & \sim & \frac{1}{2}x^2 \ln x-\frac{3}{4}x^2+\frac{x}{2} \ln(2\pi)-\frac{1}{12}\ln x \nonumber\\
	        &         & {}+ \zeta'(-1) + \widetilde{\varphi}(x),\qquad\mathrm{as}\quad x\to \infty,
	\label{id1}
\end{eqnarray}
where $\widetilde{\varphi}(x)$ is the asymptotic negative power series
\begin{equation}
	\label{sas}
	\widetilde{\varphi}(x)
	=
	\sum_{m=2}^{\infty}\frac{B_{2m}}{2m(2m-2)}\frac{1}{x^{2m-2}}.
\end{equation}
Incidentally, we mention here that the asymptotic series $\widetilde{\varphi}(x)$ provides the connection
between Penner models and string theory~\cite{GR90}, because $\widetilde{\varphi}(\rmi\mu)$  determines  
the genus expansion $\widetilde{F}_{c=1}(\mu)$ of the free energy of the $c=1$
string theory at the self-dual radius.  The sector of validity of equation~(\ref{id1}) is often quoted as $-\pi<\arg z<\pi$ or,
more precisely, $|\arg z|\leq \pi-\delta$ with $\delta >0$ (see equation~5.17.5 in~\cite{OL10}).
We want to emphasize that the expansion~(\ref{id1}) is asymptotic in this sector only in the sense of Poincar\'e.

To evaluate the $G$ function for large negative real values of the argument we use the reflection formula 
(see equation~(6) of~\cite{AD01})  to obtain
\begin{eqnarray}
	\label{bneg}
 	\ln |G(1-x)|  =  \ln G(1+x) + x \ln\left| \frac{\sin(\pi x)}{\pi} \right| + \frac{1}{2\pi}\cltwo(2\pi x),
\end{eqnarray}
where $\mathrm{Cl}_2(x)$ is the Clausen function
\begin{equation}
	\label{cla}
	\cltwo(x)  =  - \int_0^x \ln\left| 2 \sin \frac{\tau}{2} \right |\rmd\tau
	               = \sum_{m=1}^{\infty}\frac{\sin(m x)}{m^2}.
\end{equation}
Note that equation~(6) in~\cite{AD01} is limited to $0<x<1$.

Thus,  from~(\ref{id1}) and~(\ref{bneg}) we obtain
\begin{eqnarray}
	\label{nxb}
\fl	& &\ln|G(1-x)|- x \ln\left| \frac{\sin(\pi x)}{\pi} \right| -\frac{1}{2\pi} \cltwo(2\pi x)
        \sim \nonumber\\
        & & \qquad\frac{1}{2}x^2 \ln x- \frac{3}{4}x^2 +\frac{x}{2}\ln(2\pi) -\frac{1}{12}\ln x+\zeta'(-1)
        +
        \widetilde{\varphi}(x),\nonumber\\
        & & \qquad\qquad x\rightarrow\infty.  
\end{eqnarray} 
The second term in the left-hand side of~(\ref{nxb}) cancels the singularities of $\ln|G(1-x)|$ at positive integer
values of $x$, and therefore the left-hand side of~(\ref{nxb}) has a well-defined asymptotic expansion.
However, in order to have a well-defined large-$n$ free energy of the Penner model with negative coupling
constant, we need to control the behavior of $\ln|G(1-x)|$ as $x$ becomes large.
Obviously the terms $x \ln\left| \sin(\pi x)/\pi \right|$ and $\cltwo(2\pi x)/2\pi$ in~(\ref{nxb})
are the origin of the $\mathcal{F}^{(\rm osc)}$ in~(\ref{nopf}) and  $l$-dependent contributions in~(\ref{nopfp})
arising in the planar free energy of the Penner model. Therefore we have to restrict the way in which the
coupling constant  $g$ tends to zero (and consequently $x$ to infinity) in such a way
that these terms give well-defined contributions to the planar limit.
This is the ultimate reason for requiring the  existence of the limit~(\ref{eq:l}).
\section*{Appendix B. The topological expansion of the Penner model with positive coupling constant}
The large $n$ expansion of the Penner model with positive coupling constants for {}'t~Hooft sequences can be
readily obtained from equations~(\ref{ane0}) and~(\ref{id1}),
\begin{eqnarray}
	\label{top}                                                                
	F_n(t) & = &  - \frac{\ln|Z_n(g)|}{n^2} \nonumber\\
	           & \approx & -\left(\frac{(t+1)^2}{2t^2} \ln (1+t)-\frac{3}{4}-\frac{1}{2 t}\right)
	                              + \frac{1}{12 n^2}\ln (1+t) \nonumber\\ 
                   &  & {}-\sum_{k=2}^\infty \frac{B_{2k}}{2k(2k-2)} n^{-2k}t^{2k-2}\left( (1+t)^{2-2k}-1 \right)\nonumber\\
                  & = &-\sum_{k \geq 0}^{\infty}n^{-2k}\sum_{s>0, 2-2k-s<0}
                            \frac{(-1)^s (2k+s-3)! (2k-1)}{(2k)! s!}B_{2k}\,t^{2k+s-2}\nonumber\\
                   &    &  \quad n\rightarrow \infty.
\end{eqnarray}
Alternatively, the standard perturbative method applied to~(\ref{ane0}) leads to a topological  expansion
of the form~\cite{PE88,MU98}                                        
\begin{equation}
	\label{top2}
	F_n(t)\approx -\sum_{k \geq 0}^{\infty}n^{-2k}\sum_{ s>0, 2-2k-s<0}\chi_{k,s}\,t^{2k+s-2},
\end{equation}
where  $\chi_{k,s}$ is the virtual  Euler characteristic  of the space of Riemann surfaces of genus $k$
with a finite number $s$  of punctures. Then equations~(\ref{top}) and~(\ref{top2}) imply~\cite{PE88}
\begin{equation}
	\label{chii}
	\chi_{k,s} = \frac{(-1)^s(2k+s-3)!(2k-1)}{(2k)!s!}B_{2k}.
\end{equation}
\section*{Acknowledgments}
We thank Prof.~A.~Mart\'{\i}nez  Finkelshtein  for calling our attention to many nice  results on zero asymptotics
of Laguerre and Jacobi polynomials. The financial support of the Spanish Ministerio de Econom\'{\i}a y
Competitividad under Project No. FIS2015-63966-P is gratefully acknowledged.
\section*{References}
\providecommand{\newblock}{}

\end{document}